\documentclass[aps,amssymb]{revtex4}

\begin{document}

\title{Does a Nash theory of gravity make dark energy superfluous?}
\author{Kayll Lake}
\email{lakek@queensu.ca}

\affiliation{Department of Physics, Queen's University, Kingston, Ontario, Canada, K7L 3N6}

\begin{abstract}
Recently Aadne and Gr{\o}n have argued that dark energy may follow naturally from a Nash theory of gravity. In this brief note I argue why this cannot be the case.
\end{abstract}
\maketitle
The Nash tensor ($N$) \cite{nash} is a symmetric two-index tensor given by
\begin{equation}\label{orgfield}
\Box  G^{\mu \nu} + G^{\alpha \beta}(2 R_{\alpha\;\;\beta}^{\;\;\mu\;\;\nu}-\frac{1}{2} g^{\mu \nu} R_{\alpha \beta}) \equiv N^{\mu \nu},
\end{equation}
where $G_{\alpha \beta}$ is the Einstein tensor, $R_{\alpha \beta \gamma \delta}$ the Riemann tensor, $R_{\alpha \beta}$ the Ricci tensor, and $g_{\alpha \beta}$ the metric tensor. $\Box$ is the covariant d'Alembertian. It is clear that Nash hoped to develop a theory of gravity based on this tensor. Without such a theory, we restrict ourselves to the case $N^{\mu \nu} = 0$. It is convenient to write a (somewhat) generalized version of this tensor in mixed form,
\begin{equation}\label{modfield}
  N_{\nu}^{\mu} = \Box G_{\nu}^{\mu} + G_{\beta}^{\alpha}(2 \epsilon R_{\alpha\;\;\nu}^{\;\;\mu\;\;\beta}-\frac{1}{2} \delta_{\nu}^{\mu}R_{\alpha}^{\beta})
\end{equation}
where $\epsilon = \pm 1$, the original Nash tensor having $\epsilon=-1$. Nash's web site is still available at Princeton \cite{nashsite} and there one can find an entry dated 27 June 2003 which states that

\bigskip

  ``\textit{I just happened today to notice that the equation, as an equation
for a vacuum space-time, is satisfied by an Einstein space (where
the Ricci curvature tensor is in a fixed proportion to the metric
tensor) PROVIDED that the dimension is four (!).}"

\bigskip

Indeed, setting
\begin{equation}\label{vacuum}
  R_{\alpha}^{\beta} = \Lambda \delta_{\alpha}^{\beta},
\end{equation}
it follows from (\ref{modfield}) that
\begin{equation}\label{nlambda}
  N_{\nu}^{\mu} = \Lambda^2 (\frac{n}{2}+\epsilon2) \delta_{\nu}^{\mu}
\end{equation}
where $n$ is the dimension of the space. Clearly every ``pure" vacuum solution $R_{\alpha}^{\beta}=0$ has zero Nash tensor (\ref{modfield}). Further, exactly as he said, Nash's original tensor is zero for an Einstein $\Lambda$ - vacuum in four dimensions. (Note that for $\epsilon = +1$ the Nash tensor is never zero for an Einstein $\Lambda$ - vacuum unless it is a pure vacuum solution.) However, with (\ref{vacuum}), the properties of the full Nash tensor are not really tested since $\Box G_{\nu}^{\mu} = 0$ automatically. In particular, whereas (\ref{vacuum}) $\Rightarrow N_{\nu}^{\mu} = 0$ for $n=4$ and $\epsilon = -1$, the problem of exactly what $g_{\alpha \beta}$ give $N_{\nu}^{\mu} = 0$, is rather more difficult.

\bigskip

Recently, Aadne and Gr{\o}n \cite{Gron} have attempted a solution to this problem in a restricted static spherically symmetric case \cite{units}
\begin{equation}\label{ssstatic}
  ds^2 = -f(r)dt^2+\frac{dr^2}{f(r)}+r^2(d\theta^2+\sin(\theta)^2d\phi^2).
\end{equation}
They found that for (\ref{ssstatic}) $N_{\nu}^{\mu} = 0$ (with  $\epsilon = -1$) for
\begin{equation}\label{schds}
  f(r)=1-\frac{2m}{r}-\frac{\Lambda r^2}{3},
\end{equation}
that is, the Schwarzschild - de Sitter (Kottler) metric. However, it is not even necessary to calculate the Nash tensor in this case. Solving $R_{\alpha}^{\beta} = \Lambda \delta_{\alpha}^{\beta}$ for (\ref{ssstatic}) gives (\ref{schds}) and so we already know that the Nash tensor vanishes due to (\ref{nlambda}). Further, Aadne and Gr{\o}n \cite{Gron} examine the spatially flat Robertson - Walker metric
\begin{equation}\label{rw}
  ds^2=-dt^2+a(t)(dr^2+ r^2(d\theta^2+\sin(\theta)^2d\phi^2))
\end{equation}
and find that $N_{\nu}^{\mu} = 0$ (with  $\epsilon = -1$) for
\begin{equation}\label{ds}
  a(t)=\exp{\sqrt{\frac{\Lambda}{3}}t}.
\end{equation}
However, once again, it is not necessary to calculate the Nash tensor. Solving $R_{\alpha}^{\beta} = \Lambda \delta_{\alpha}^{\beta}$ for (\ref{rw}) gives (\ref{ds}) and so we already know that the Nash tensor (with $\epsilon=-1$) vanishes due to (\ref{nlambda}). (Indeed, this case is a coordinate transformation of the previous case with $m=0$.) The suggestion in \cite{Gron} is that $\Lambda$ develops naturally from (\ref{modfield}) (for $n=4$ and $\epsilon = -1$). However, there is no evidence for this. Of the two examples presented, $\Lambda$ does not develop naturally from (\ref{modfield}) (with $\epsilon = -1$), but rather it develops from the fact that (\ref{vacuum}) is satisfied. Moreover, since every ``pure" vacuum has $\Lambda = N_{\nu}^{\mu} = 0$, it is certainly not clear how $N_{\nu}^{\mu} = 0$ can generate $\Lambda$.

\bigskip

A useful example at this point would be a case for which $ N_{\nu}^{\mu} = 0$ for $n=4$ and $\epsilon = -1$ with $\Box G_{\nu}^{\mu} \neq 0$ and  $R_{\alpha}^{\beta} \neq \Lambda \delta_{\alpha}^{\beta}$ so that the Nash tensor \textit{has} to be calculated. It is not difficult to find such an example. Any conformally flat spacetime with conformal factor
\begin{equation}\label{conformal}
 (\Lambda x^2+2(\Lambda c)^{1/2}x+c)(\lambda y^2+2(\lambda d)^{1/2}y+d)(\delta z^2+2(\delta e)^{1/2}z+e)
\end{equation}
provides such an example (I have used \textit{GRTensor II} with \textit{Maple} \cite{grt}).
Here $x, y$ and $z$ are spatial coordinates, and the coefficients $\Lambda, \lambda$ and $\delta$ can bet set to $1$ by choice of scale. Also $c, d$ and $e$ are constants.

\bigskip

\textit{Acknowledgments.} This work was supported by a grant from the Natural Sciences and Engineering Research Council of Canada. It is a pleasure to thank Eric Poisson for discussions.


\begin{thebibliography}{}\label{sec:TeXbooks}
\bibitem{nash}Lecture by John F. Nash Jr. ``\textit{An Interesting Equation.}"
\texttt{http://sites.stat.psu.edu/~babu/nash/intereq.pdf}
\bibitem{nashsite}\texttt{http:/web.math.princeton.edu/jfnj/}
\bibitem{Gron}M. T. Aadne and O. G. Gr{\o}n, ``\textit{Exact Solutions of the Field Equations for Empty Space in the Nash Gravitational Theory}", Universe 2017, 3(1), 10    [arXiv:1702.06833].
\bibitem{units}I use geometrical units throughout.
\bibitem{grt} This is a package which runs within Maple. It is entirely distinct from packages distributed with Maple and must be obtained independently. The \textit{GRTensorII} software and documentation is distributed freely on the World-Wide-Web from the address \texttt{http://grtensor.org}.  \textit{GRTenorIII}, developed by Peter Musgrave, is now available free of charge. Release information is at:

\texttt{http://hyperspace.uni-frankfurt.de/2016/12/07/grtensoriii-for-maple-has-been-released/}

and access is at:

\texttt{https://github.com/grtensor/grtensor}
\end{thebibliography}
\end{document}